\documentclass[usenatbib,letterpaper]{mn2e}
\usepackage{graphicx}
\usepackage{rotating}
\usepackage{blindtext}
\usepackage{amstext}
\usepackage{url}

\title[Magnetic field Evolution]
  {On the diversity of compact objects within supernova remnants -- I. A parametric model for magnetic field evolution}
\author[Adam Rogers and Samar Safi-Harb]
{Adam Rogers\thanks{E--mail: rogers@physics.umanitoba.ca} and Samar Safi-Harb\thanks{Canada Research Chair. E--mail: samar.safi-harb@umanitoba.ca} \\ Department of Physics and Astronomy, University of Manitoba, Winnipeg, Manitoba, R3T 2N2, Canada}

\date{Accepted 2015 December 28. Received 2015 December 11; in original form 2015 July 10}
\volume{457}
\pagerange{1180--1189} \pubyear{2016}

\begin{document}
\label{firstpage}
\maketitle

\begin{abstract}
A wealth of X-ray and radio observations has revealed in the past decade a growing diversity of neutron stars (NSs) with properties spanning orders of magnitude in magnetic field strength and ages, and with emission processes explained by a range of mechanisms dictating their radiation properties. However, serious difficulties exist with the magneto-dipole model of isolated NS fields and their inferred ages, such as a large range of observed braking indices ($n$, with values often $<$3) and a mismatch between the NS and associated supernova remnant (SNR) ages. This problem arises primarily from the assumptions of a constant magnetic field with $n$ = 3, and an initial spin period that is much smaller than the observed current period. It has been suggested that a solution to this problem involves magnetic field evolution, with some NSs having magnetic fields buried within the crust by accretion of fall-back supernova material following their birth. In this work, we explore a parametric phenomenological model for magnetic field growth that generalizes previous suggested field evolution functions, and apply it to a variety of NSs with both secure SNR associations and known ages. We explore the flexibility of the model by recovering the results of previous work on buried magnetic fields in young NSs. Our model fits suggest that apparently disparate classes of NSs may be related to one another through the time evolution of the magnetic field.
\end{abstract}

\begin{keywords}
stars: magnetars - stars: magnetic field - stars: neutron - stars: rotation - ISM: supernova remnants - X-rays: stars
\end{keywords}

\section{Introduction}
\label{sec:intro}
Thanks to the advance of modern X-ray telescopes such as \textit{Chandra} and \textit{XMM-Newton}, and the synergy with radio observations, we now know that isolated neutron stars (NSs) can manifest themselves as pulsars (PSRs) with a surface dipole magnetic field spanning more than five orders of magnitude, in the $\sim$10$^{10}$--10$^{15}$~G range\protect{\footnote{For the remainder of this work we use the equatorial surface dipole field, $B=3.2 \times 10^{19}\sqrt{ P \dot{P} }$~(G), inferred from the observed period, $P$(s), and period derivative, $\dot{P}$ (s s$^{-1}$).}}. Observationally, this has led to their organization into different classes, including (1) the rotationally powered radio and X-ray bright objects, like the Vela pulsar with $B$$\sim$$10^{11}$--$10^{13}$~G, (2) the magnetically powered pulsars (or magnetars) with $B$$\sim$$10^{14}$--$10^{15}$~G, exceeding the quantum electrodynamics (QED) limit of $4.4 \times 10^{13}$ G and observed primarily at high energies, (3) the highly magnetized pulsars (HBPs) with magnetic fields intermediate between the classical pulsars and magnetars, but still exceeding the QED limit, and (4) the central compact objects (CCOs) observed only in X-rays (so far), near the centres of supernova remnants (SNRs) and with inferred low magnetic fields, $B$$\sim$10$^{10}$--10$^{11}$~G. This diversity led several authors to attempt a unification through evolutionary models of NSs with their properties dictated primarily by a continuum of magnetic field strengths \citep[see e.g.][and references therein]{kaspi2010, dallOsso, mereghetti13, perna2013, vigano2013, safiharb15}.

The magnetic field is estimated using a standard model of NS evolution which assumes energy loss due to the emission of radiation from a point-like rotating magnetic dipole in vacuum, providing a spin-down torque with a braking index $n=3$ \citep{dipole}. This picture assumes rapid rotation of the NS after birth, so the observed period ($P$) differs from the initial period ($P_0$) by a large amount (i.e. $P_0 \ll P$) due to the constant torque acting to slow the NS spin \citep{spin}. However, the braking index has been measured for a small sample of young pulsars, and all so far differ from the prediction of the standard model with $n<3$ \citep{youngn}. A lower braking index can come about from a variety of mechanisms including a change in the moment of inertia of the star over time \citep{moi}, alignment of the magnetic field and rotation axis \citep{alignment1, alignment2}, the emission of a particle wind \citep{wind1, wind2}, magnetospheric effects \citep{magS1, magS2} and environmental interactions \citep{fallback}. Another serious problem with the standard picture concerns the NSs that are associated with SNRs. Generally, pulsar ages found from their `characteristic age' ($\tau_\text{PSR}=P/2 \dot{P}$) by assuming dipole radiation and the independently measured SNR ages are in disagreement, sometimes by orders of magnitude (in particular for the CCOs)\footnote{See http://www.physics.umanitoba.ca/snr/SNRcat for the high-energy catalogue of SNRs which compiles all known ages of SNRs and associated PSRs \citep{SNRCatRef}.}. The observed braking index and age discrepancy arise from the standard, and commonly adopted, assumption of a constant torque acting to brake the NS over its life span.

The growing evidence for NSs with X-ray luminosity in excess of their spin-down energy presents another difficulty for the standard scenario. Some of these objects are thought to be powered by the dissipation of magnetic energy rather than spin-down losses, examples of which include the anomalous X-ray pulsars (AXPs) and some soft gamma-ray repeaters (SGRs), unified under the class of `magnetars' \citep{magnetar1, magnetar3, magnetar2}. These are X-ray bright objects that are slowly rotating pulsars with exceptionally high magnetic fields and normally discovered through their bursting activity. However, a neat classification scheme for these objects proves elusive in the light of the discovery of `low-B magnetars' \citep[e.g.][]{rea2010}, and an HBP having behaved like a magnetar, yet thought to be a rotation-powered pulsar powering a bright pulsar wind nebula \citep{HBP2, HBP1}. The situation is further complicated by the CCOs with extremely low fields, dubbed as `anti-magnetars' \citep{CCO1}, yet still show an X-ray luminosity in excess of their spin-down energy. One recent interpretation for these objects is the suppression of their external field through magnetic field burial \citep{burial0, burial1, burial2, burial3}, also implied by spectroscopic models of these objects \citep{CCO2}. In this scenario, the accretion of supernova fall-back material occurs following the birth of the NS. This period of vigorous accretion has the effect of burying the dipole magnetic field component within the NS crust, reducing the spin-down energy loss and making the NS appear significantly older than its associated SNR. In this alternative model of NS evolution, field growth is needed to explain the initially small braking index and low surface fields, while a decaying toroidal component is invoked to explain the excess X-ray luminosity \citep{HoCCOReview}. The field burial scenario has been most recently described in significant detail by \citet{ho15}, who performed detailed calculations of the fall-back accretion process, including the inner structure of the NS, conductivity of the NS crust and a realistic equation of state. Besides an internal decaying toroidal component, the dipole field component also decays on large time-scales \citep{dallOsso, vigano2013}. The decaying external field is described by a parametrized model given by \citet{colpi}, expanded on by \citet{dallOsso} and used to describe the evolution of the AXP 1E~2259+586 by \citet{nakano15}.

In this paper we present a phenomenological parametrized family of models for magnetic field evolution in the NS population. Our model unifies the description of magnetic field growth and decay by making use of variations on the parametric forms from \citet{dallOsso} and \citet{colpi}, which we derive in Section \ref{sec:theory}. This model also reproduces the results of \citet{NB15} for exponential field growth and replicates the findings of \citet{dallOsso} for decaying fields. We fit our model to the observations of various NSs in Section \ref{sec:modelfitting}, testing our model against the detailed physical predictions found by \citet{ho15}. We discuss the results of our fits in Section \ref{sec:discussion}. Finally, our conclusions are summarized in Section \ref{sec:conclusions}.

\section{Theory and Parameter Space Exploration}
\label{sec:theory}
The standard model for NS spin-down from energy loss due to the emission of dipole radiation assumes a constant magnetic field $B \propto \sqrt{P \dot{P}}$, where $P$ and $\dot{P}$ are the period and period derivative, respectively. However, we are interested in the dynamical evolution of the magnetic field
\begin{equation}
B(t)=B_{\text{j}} f_{\text{j}}(t),
\label{Bt}
\end{equation}
where the time-dependence has been gathered into the function $f_{\text{j}}(t)$ \citep[see also ][]{IP_2} and $B_{\text{j}}$ is a constant reference field value, either the initial field strength in decay models (denoted by subscript $D$) or final field strength in growth models (labelled as $G$). We use the differential equation
\begin{equation}
P \dot{P} = b B^2,
\label{funDip}
\end{equation}
with $b=\text{constant}$, and do not consider the effect of spin-axis field alignment. Integrating this equation from the NSs birth at $t=0$ to an arbitrary later time $t$, we find
\begin{equation}
P^2=P_0^2+2bB_{\text{j}}^2 F_{\text{j}}^2
\label{P}
\end{equation}
where we have denoted the integral
\begin{equation}
F_{\text{j}}^2=\int_0^t f_{\text{j}}^2(t') dt'.
\label{F_j^2}
\end{equation}
From here on, we will generally suppress the time-dependence of the function $f$ for notational simplicity. The period derivative is
\begin{equation}
\dot{P}=\frac{bB_{\text{j}}^2f_{\text{j}}^2}{P},
\label{dotP}
\end{equation}
and we express the characteristic age, $\tau$, as
\begin{equation}
\tau=\frac{P}{2\dot{P}}=\frac{P^2}{2bB_{\text{j}}^2f_{\text{j}}^2},
\label{ttau1}
\end{equation}
where we have used equation (\ref{dotP}). Inserting equation (\ref{P}) in this expression gives us
\begin{equation}
\tau=\frac{P_0^2}{2bB_{\text{j}}^2f_{\text{j}}^2}+\frac{F_{\text{j}}^2}{f_{\text{j}}^2}.
\label{tau}
\end{equation}
Including a time-dependent field introduces an important distinction between the characteristic age $\tau$ and the model time $t$. The model time represents the amount of time elapsed from the birth of the NS to the present, and thus represents the ``true'' age of the NS. The characteristic age is the age that is determined from the period and period derivative of the NS (equation \ref{ttau1}), which differs from the true age due to the time-dependence of $P$ and $\dot{P}$. These dynamical quantities also introduce a time-dependence of the braking index $n=2-P\dot{P}/\ddot{P}$. Using equation (\ref{P}), the braking index is given by
\begin{equation}
n= 3 - 4 \tau \frac{\dot{f_{\text{j}}}}{f_{\text{j}}}.
\label{n}
\end{equation}
If the field decays, $\dot{f_{\text{j}}}$ is negative and $n>3$, while field growth has positive $\dot{f_{\text{j}}}$ and leads to a braking index $n<3$ as observed in many young NSs. In the case of $f_{\text{j}}=1$, $F_{\text{j}}^2=t$, and these formulae reduce to the standard spin-down from dipole radiation with a constant field.

\citet{dallOsso} use a parametrization to study field decay that was introduced by \citet{colpi}:
\begin{equation}
\frac{dB}{dt}=-aB(t)^{1+\alpha}=-\frac{B(t)}{\tau_{\text{D}}}
\end{equation}
where $\alpha$ is the decay index that describes how rapidly the decay proceeds and $a$ is a normalization parameter related to the specific physical mechanisms involved in the decay (e.g. Hall drift, ambipolar diffusion). The quantity $\tau_{\text{D}}=\left[ aB(t)^{\alpha} \right]^{-1}$ is the decay time-scale, which itself is time-dependent. The magnetic field described by these equations can be conveniently written as
\begin{equation}
B(t) = B_{\text{D}} \left\{
\begin{array}{ll}
\left( 1+ \alpha \frac{t} {\tau_{\text{m}}} \right)^{-\frac{1}{\alpha}} , & \alpha\neq0,2 \\
\exp\left( -\frac{t}{\tau_{\text{m}}} \right), & \alpha=0 \\
\end{array}\right.
\label{decayB}
\end{equation}
where $B_{\text{D}}$ is the initial field which decays over time, and $\tau_{\text{m}}=\tau_{\text{D}}(0)$ is the initial field decay time-scale. Note that the time-dependence of the decaying field given in equation (\ref{decayB}), $B(t)=B_{\text{D}} f_{\text{D}}(t)$, is contained entirely within the dimensionless function $f_\text{D}(t)$. This well-known parametrization is extremely useful as it can also be used to construct a model of field growth.

There are a number of properties the dimensionless function $f_\text{G}(t)$ must have in order to describe NS field growth. The growth must be bounded in time, in that the field begins at some minimum value and attains an asymptotic maximum value as time increases. This requires that the derivative of the field growth function is always positive and decreases to approach zero as $t$ becomes large. Furthermore, $f_\text{G}(t)$ should be parametrized in terms of a small number of quantities whose meaning has a clear physical interpretation. In fact, the field decay function $f_\text{D}(t)$ has attractive features that make it useful to also describe the derivative of a growing field $\text{d}{f_\text{G}}/\text{dt}$. First, the function decreases from a maximum $f_\text{D}(0)=1$ and becomes vanishingly small for large $t$. This bounded behaviour fulfills the exact criteria that is required to describe the derivative of a field $\text{d}{f_\text{G}}/\text{dt}$ that begins at a minimum value and grows to approach a constant strength as $t$ increases. Secondly, $f_\text{D}(t)$ is stated in terms of two parameters that have a well-understood interpretation. The index $\alpha$ controls the rate at which $f_\text{D}(t)$ changes with respect to $t$, with lower values giving the field evolution an exponential behaviour, and larger values slow the evolution providing a softer decay. The parameter $\tau_\text{m}$ controls the time-scale over which the magnetic field evolves. Therefore, let us consider the following basic form based on the field decay $f_\text{D}(t)$ from \citet{dallOsso}, with an appropriate normalization, as
\begin{equation}
\frac{\text{d}f_{\text{G}}}{\text{d}t}=\frac{(1-\alpha)}{\tau_{\text{m}}} f_{\text{D}}
\label{Bderiv}
\end{equation}
where $f_{\text{G}}$ is the time-dependent part of the growing field and $f_{\text{D}}$ contains the time-dependence of the decaying field model, normalized by the factor $(1-\alpha)/\tau_\text{m}$. Due to this normalization, growing fields require the field index to be in the range $0 \leq \alpha <1$. Equation \ref{Bderiv} results in a field that evolves in time as $B(t)=B_{\text{G}}f_{\text{G}}(t)$, where
\begin{equation}
f_{\text{G}}(t) = \epsilon + \left\{
\begin{array}{ll}
1-\left( 1+\alpha \frac{t}{\tau_{\text{m}}} \right)^{\frac{\alpha-1}{\alpha}}, & 0 < \alpha < 1 \\
1-\exp\left(-\frac{t}{\tau_{\text{m}}}\right), & \alpha=0
\end{array}\right. .
\label{fG}
\end{equation}
In the above expression the integration constant $\epsilon$ controls the initial field
\begin{equation}
B_0=B_{\text{G}} \epsilon
\label{BInit}
\end{equation}
in terms of the asymptotic value at large $t$, $B_{\text{G}}$. The growth model uses the boundary condition $B_{\text{j}}=B_{\text{G}}$ as the asymptotic field strength, and the decay model uses $B_{\text{j}}=B_{\text{D}}$ as the initial field. In terms of fields buried by fall-back accretion, the smaller $\epsilon$ is, the deeper the field has been buried within the NS, and the larger the difference between the initial and asymptotic field strength. The time-scale $\tau_{\text{m}}$ determines how long the field takes to emerge from the compact object. The field given by equation (\ref{fG}) describes a family of solutions in terms of the field index $\alpha$. When $\alpha=0$ the field evolves exponentially, which is particularly significant as this form was proposed by \citet{NB15} to describe growing NS fields. When $0<\alpha<1$, field growth occurs more slowly. Since the parameters of our growth model are given by the widely studied decaying field parametrization of \citet{dallOsso} and \citet{colpi}, we find this representation to be particularly intuitive.

The period $P$ and characteristic age $\tau$ given by equations (\ref{P}) and (\ref{tau}) depend on the integral of $f_{\text{G}}^2$ (equation \ref{F_j^2}). With the time-dependence from equation (\ref{fG}), we write
\begin{equation}
\begin{array}{ll}
F_{\text{G}}^2 =\int_0^t f_{\text{G}}^2(t')dt' = & (1+\epsilon)^2t \\
  &   -\frac{2(1+\epsilon)\tau_{\text{m}}}{2 \alpha-1}\left[ \left( 1 + \alpha \frac{t}{\tau_{\text{m}}}\right)^{\frac{2 \alpha-1}{\alpha}}-1\right] \\
  &   +\frac{\tau_{\text{m}}}{3 \alpha-2} \left[ \left( 1+\alpha \frac{ t}{\tau_{\text{m}}}\right)^{\frac{3\alpha-2}{\alpha}}-1\right].
\end{array}
\label{FG2}
\end{equation}
We give the special cases of this equation in the limits $\alpha \rightarrow 1/2$ and $\alpha \rightarrow 2/3$, and summarize the connection between the field decay and growth models in Table \ref{table1}.

\setcounter{table}{0}
\begin{table*}
\begin{center}
\begin{tabular}{|c|c|c|}

\multicolumn{3}{c}{Decay functions} \\ \hline

$\frac{\text{d}f_{\text{D}}}{\text{d}t}$ & $-\frac{1}{\tau_{\text{m}}}\left( 1+\alpha \frac{t}{\tau_{\text{m}}} \right)^{-\frac{(\alpha+1)}{\alpha}}$     & $0 \leq \alpha \leq 2$ \\
$$ & $-\frac{1}{\tau_{\text{m}}} \exp\left( -\frac{t}{\tau_{\text{m}}} \right)$ & $\alpha=0$ \\ \hline

$f_{\text{D}}$ & $\left(1+\alpha \frac{t}{\tau_{\text{m}}} \right)^{-\frac{1}{\alpha}} $ & $$ \\
   $$   & $\exp\left( -\frac{t}{\tau_{\text{m}}} \right) $ & $\alpha = 0$ \\ \hline

$F_{\text{D}}^2=\int_0^t f_{\text{D}}^2(t')dt'$      & $\frac{\tau_{\text{m}}}{2-\alpha}\left[1-\left(1+\alpha \frac{t}{\tau_{\text{m}}} \right)^{\frac{\alpha-2}{\alpha}}\right]$ & $$ \\
$$ & $\frac{\tau_{\text{m}}}{2}\left[1-\exp\left( -2\frac{t}{\tau_{\text{m}}} \right)\right]$ & $\alpha=0$ \\
$$ & $\frac{\tau_{\text{m}}}{2}\ln \left( 1+2\frac{t}{\tau_{\text{m}}} \right)$ & $\alpha=2$ \\ \hline

\multicolumn{3}{c}{Growth functions} \\ \hline

$\frac{\text{d}f_{\text{G}}}{\text{d}t}$      & $\frac{(1-\alpha)}{\tau_{\text{m}}}\left( 1+ \alpha\frac{t}{\tau_{\text{m}}}\right)^{-\frac{1}{\alpha}}$ & $0 \leq \alpha < 1$\\
$$                     & $\frac{1}{\tau_{\text{m}}}\exp\left( -\frac{t}{\tau_{\text{m}}} \right) $ & $\alpha = 0$ \\ \hline

$f_{\text{G}}$ & $1+\epsilon -\left( 1+ \alpha \frac{t}{\tau_{\text{m}}} \right)^{\frac{\alpha-1}{\alpha}}$ & $0<\alpha<1$ \\
$$  & $1+\epsilon -\exp\left( -\frac{t}{\tau_{\text{m}}} \right)$ & $\alpha = 0$ \\ \hline

$F_{\text{G}}^2=\int_0^t f_{\text{G}}^2(t')dt'$      & $(1+\epsilon)^2t-\frac{2(1+\epsilon)\tau_{\text{m}}}{2 \alpha-1}\left[ \left( 1 + \alpha \frac{t}{\tau_{\text{m}}}\right)^{\frac{2 \alpha-1}{\alpha}}-1\right] + \frac{\tau_{\text{m}}}{3 \alpha-2} \left[ \left( 1+\alpha \frac{ t}{\tau_{\text{m}}}\right)^{\frac{3\alpha-2}{\alpha}}-1\right]$ & $$ \\

$ \lim_{\alpha \rightarrow 1/2} F_{\text{G}}^2$ & $(1+\epsilon)^2t - 4(1+\epsilon)\tau_{\text{m}} \ln\left( 1+\frac{1}{2}\frac{t}{\tau_{\text{m}}}\right)+t\left( 1+\frac{1}{2} \frac{t}{\tau_{\text{m}}}\right)^{-1}$ & $\alpha = \frac{1}{2}$ \\

$ \lim_{\alpha \rightarrow 2/3} F_{\text{G}}^2$ & $(1+\epsilon)^2t -6(1+\epsilon)\tau_{\text{m}}\left[ \left(1+\frac{2}{3}\frac{t}{\tau_{\text{m}}}\right)^{\frac{1}{2}} -1 \right]+ \frac{3}{2} \tau_{\text{m}} \ln\left( 1+\frac{2}{3}\frac{t}{\tau_{\text{m}}} \right)$ & $\alpha = \frac{2}{3}$ \\ \hline

\end{tabular}
\caption{A summary of the time-dependent functions for describing magnetic field decay and growth.}
\label{table1}
\end{center}
\end{table*}

The model as stated has six parameters: the field index $\alpha$, growth factor $\epsilon$, time-scale $\tau_{\text{m}}$, asymptotic field $B_\text{G}$, the initial period $P_\text{0}$ and the model time $t$. The model time can be treated as a free parameter to vary between the lower and upper SNR age limits, $\tau_\text{SNR--}$ and $\tau_\text{SNR+}$, respectively, or can be fixed before beginning the optimization. The model then outputs the quantities we want to fit to the observed values: the period, period derivative and braking index (at the present time). The standard fitting problem is to vary the input parameters to produce a match with the output and the observed values. Thus, the problem is under constrained, in that there are fewer fit quantities than parameters, leading to a family of solutions. However, a closer inspection shows that the previously stated input parameters are not truly independent. In fact, a simplification can be achieved by changing our modelling approach.

Instead of fitting for $P$ and $\dot{P}$, let us assume their observed values a priori. We then know the magnetic field at the present time, which we call $B_\text{t}$, by equation (\ref{funDip}), and also by definition, the characteristic age $\tau=\tau_\text{PSR}$. With the definition of the growing field in equation \ref{Bt}, we can solve for the model time $t$ as a function of the present-day field $B_\text{t}$ and the four parameters ($\alpha$, $\epsilon$, $\tau_\text{m}$, $B_\text{G}$):
\begin{equation}
t = \left\{
\begin{array}{ll}
\frac{ \tau_\text{m} }{ \alpha }\left[ \left( 1+\epsilon -\frac{B_\text{t}}{B_\text{G}} \right)^\frac{\alpha}{\alpha-1} - 1 \right], & 0 < \alpha < 1 \\
-\tau_\text{m} \ln \left( 1+\epsilon -\frac{B_\text{t}}{B_\text{G}} \right), & \alpha=0
\end{array}\right. .
\label{tModel}
\end{equation}
Once we have calculated $t$, we find $f$ and $\dot{f}$ using equation (\ref{fG}), and obtain the braking index $n$ at the current time through equation (\ref{n}). The parameters are then varied to match $n$ and $t$ to the observed values. Knowledge of the characteristic age allows us to state the initial period as a function of the parameters, given by solving equation (\ref{tau}):
\begin{equation}
P_\text{0} = \sqrt{ 2 b B_\text{G}^2 \left( \tau_\text{PSR} f_\text{G}^2 - F_{G}^2 \right) }.
\label{P0eq}
\end{equation}
Restating the problem in this way is advantageous because it allows us to eliminate what was previously considered a free parameter, and treats the SNR age and braking index as quantities to fit. The simplification we introduce comes from reducing the number of parameters to a small enough set that a quantitative description of the model parameter space can be given, as described furthermore below. The introduction of additional physics beyond the phenomenological field growth also helps further simplify the situation.

We describe the model parameter space by fixing the values of the field index $\alpha$ and the growth factor $\epsilon$. For a given calculation we hold these values constant. Next, we form a grid of $\tau_\text{m}$ and $B_\text{G}$ values and calculate $t$ and $n$ for each ($\tau_\text{m}$, $B_\text{G}$) pair using equations (\ref{tModel}) and (\ref{n}). The regions of parameter space containing solutions with $t$ and $n$ within the observed limits are found by contouring these $2D$ functions to find the level curves corresponding to $\tau_\text{SNR+}$ and $\tau_\text{SNR--}$, and the measured limits on $n$. Solutions that satisfy the constraints live in the regions between these level curves. Solutions in the region where the sets intersect satisfy both of the constraints simultaneously. Changing the values of $\alpha$ and $\epsilon$ affects the morphology of the intersection regions. Using this approach, we study the regions of the parameter space that give physically realistic solutions without the need for an external optimization routine. For the remainder of this study, we will use this contouring approach for a variety of field index in the range $0.1 \leq \alpha \leq 0.9$ and growth factor $0.001 \leq \epsilon \leq 0.1$. In practice, we find that solutions with $\epsilon<10^{-3}$ do not significantly vary from one another for a given $\alpha$, so we do not consider any $\epsilon$ lower than this. Moreover, the largest asymptotic fields also typically correspond to small $\epsilon$ for a given $\alpha$, so we choose $\epsilon=0.1$ as an upper limit that still allows a significant field growth, though in general all $\epsilon < 1$ can be used. As a final point, we note that equation (\ref{P0eq}) can produce unphysical complex valued $P_\text{0}$. Thus, we impose a further constraint from the initial period:
\begin{equation}
y=\tau_\text{PSR} f_\text{G}^2 -F_{G}^2 \geq 0,
\end{equation}
with the equality corresponding to the limit $P_\text{0}=0$. This provides a boundary between the physical and unphysical solutions in the parameter space. Therefore along with $t$ and $n$, we also produce the corresponding $y$ on the ($\tau_\text{m}$, $B_\text{G}$) grid, and find the level curve $y=0$ using a contouring method. The unphysical region can then also be excluded by intersection.

\begin{center}
\begin{figure*}
\centerline{\includegraphics[scale=0.85, bb= 92 292 509 502, clip=true]{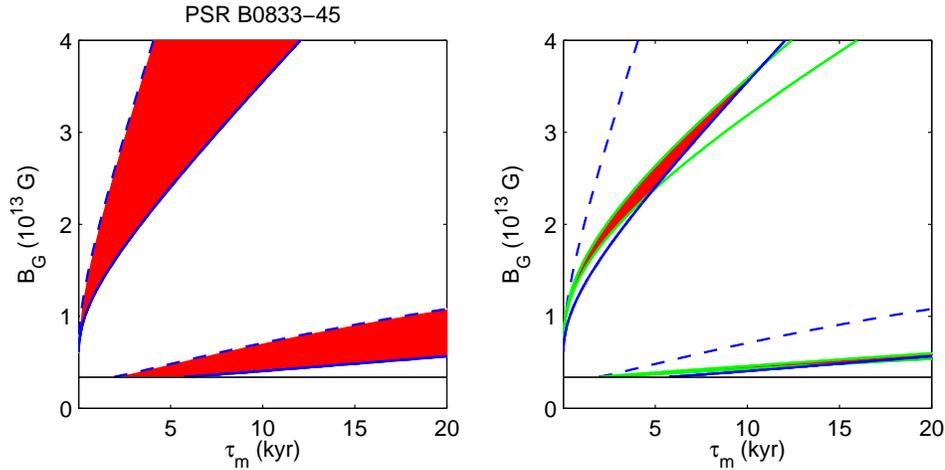}}
\caption{Parameter constraint plots for the Vela pulsar, PSR B0833--45. The left-hand panel shows the parameter space region that satisfies the age constraint (red). The lower SNR age limit is the blue dashed line, and the solid blue line is the upper SNR age limit. On the right the same region of parameter space is shown but we include the braking index constraints (green curves). The red region shows the intersecting set, which satisfies both the age constraint and the braking index constraint. The lower group in both panels has $\alpha=0.1$, $\epsilon=0.1$, and the upper group $\alpha=0.9$, $\epsilon=0.001$. The horizontal black line in both panels marks the observed dipole field.}
\label{figContours}
\end{figure*}
\end{center}

As an example, in Fig. \ref{figContours} we show the parameter space for the Vela pulsar, PSR~B0833--45, which has a measured braking index and associated SNR age. On the left, we show the ($\tau_\text{m}$, $B_\text{G}$) parameter space using only the age constraint from the associated SNR. The lower limit $\tau_{\text{SNR}-}$ is the dashed blue line, and the upper limit $\tau_{\text{SNR}+}$ is the solid blue line. The region between the curves is the parameter space area that obeys the age constraint, and is coloured red. On the right we show the same region of parameter space, but include the braking index contours, denoted as green lines. Since the braking index is known to high accuracy, the red coloured regions satisfying both constraints simultaneously is significantly reduced. The lower group in both panels has $\alpha=0.1$, $\epsilon=0.1$, and the upper group $\alpha=0.9$, $\epsilon=0.001$. Note that there is degeneracy between the groups of solutions for a given $\tau_\text{m}$ depending on the chosen field index $\alpha$. Thus, while this method does not provide a unique solution, it allows us to quantify the regions of parameter space that contain physical solutions given a constant $\alpha$ and $\epsilon$ pair as input. We investigate the differences between the low- and high-$\alpha$ cases to study the limiting behaviour of the field growth model. Generally, the solutions that have observed fields close to the asymptotic field $B_\text{G}$ are near the end of their evolution. The $\alpha=0.1$, $\epsilon=0.1$ solutions with the lowest asymptotic fields have nearly finished their evolution and will grow by only $\approx 1 \%$ over the next few kyr. The braking index of these solutions will rapidly grow to the dipole value $n=3$ over this span of time. The solutions with $\alpha=0.9$ and $\epsilon=0.001$ have significantly larger asymptotic fields that are more than an order of magnitude higher than the low-$\alpha$ solutions. These NSs will take a significantly longer span of time for their fields to evolve to the final state and reach braking index $n=3$. For these solutions, the field growth significantly outlasts the observable life of the SNR and will appear as a highly magnetized, isolated NS with no apparent SNR association.

\section{Model Fitting and NS Evolution}
\label{sec:modelfitting}

Let us investigate the consequences of field growth in NSs using two initial approaches. First, we vary the field index $\alpha$ to demonstrate how this parameter affects the time evolution. Secondly, since this field growth model is phenomenological in nature, we investigate how well it can reproduce the results of numerical simulations, such as the detailed modelling of the burial and emergence of the magnetic fields in young accreting NSs explored recently by \citet{ho15}. That study focused on the young NSs with braking indices $n<2$, in particular the rotation-powered pulsars PSR J0537--6910 associated with the LMC SNR~N157B, the Vela pulsar B0833--45, and the HBP~J1734--3333 which has a proposed association with G354.8--0.8 \citep{manchester}. We note that the relationship between this SNR and HBP~J1734--3333 is tenuous and may be the result of a coincidental alignment. We list these systems in Table \ref{table2}, along with a carefully selected list of other NSs that are (1) securely associated with SNRs (thus providing an independent estimate of the true age) and (2) with `extremal' fields, namely from the class of magnetars, HBPs and CCOs.

\begin{table*}
\small
\begin{center}
\begin{tabular}{|c|c|c|c|c|c|c|c|}
\multicolumn{8}{c}{Observed properties of NSs} \\
\hline

PSR & $P$ & $\dot{P}$                  & $n$   &  $\tau_\text{PSR}$ & SNR & $\tau_\text{SNR--}$ & $\tau_\text{SNR+}$ \\
        & (s)     & ($10^{-11}s~s^{-1}$) &           &                   (kyr) &         & (kyr)                   & (kyr) \\ \hline
\hline

AXP~1E~1841--045 & $11.783$ & $3.930$    & $$ & 4.750 &  G27.4+0.0 (Kes~73)  & $0.750$  & $2.100 $ [1]  \\
AXP~1E~2259+586 &  $6.979$  & $4.843e-2$ & $$ & 228.317 & G109.1-01.0  (CTB~109)& $10.000$ & $16.000$ [2] \\
CXOU~J171405.7--381031 & $3.825$  & $6.400$ & $$ & 0.947 & G348.7+00.3 & $0.350$ & $3.150$ [3] \\
\hline

SGR~0526--66 & $8.054$ & $3.800$ & $$ & $3.358$ & N49 & $-$ & $4.800$ [4] \\
SGR~1627--41 & $2.595$ & $1.900$ & $$ & $2.164$ & G337.3--0.1 & $-$ & $5.000$ [5]  \\
\hline

HBP~J1119--6127   & $0.408$ & $0.400$ & $2.684\pm 0.002$ {[14]}& 1.616 & G292.2--0.5 & $4.200$ & $7.100$ [6] \\
HBP~J1734--3333   & $1.170$  & $0.228$ & $ 0.9 \pm 0.2$ {[15]}  & 8.131 & G354.8--0.8 & $1.300$ & $-$ [7] \\
HBP~J1846--0258 A & $0.325$ & $0.709$ & $2.64 \pm 0.01$ {[16]} & $0.726$ & G029.7--0.3 (Kes~75)& $0.900$ & $4.300$ [8] \\
HBP~J1846--0258 B & $0.327$ & $0.711$ & $2.16 \pm 0.13$ {[17]} & $0.728$ &                      &         & \\
\hline

PSR~J0537--6910 & $0.016$ & $0.518$ & $-1.5 \pm 0.1$ {[18]} & $4.925$ & N157B & $1.000$ & $5.000$ [9] \\
PSR~B0833--45     & $0.089$ & $1.250$ & $1.4 \pm 0.2$ {[19]}& $11.319$ & G263.9--03.3 (Vela) & $5.400$ & $16.000$ [10] \\ \hline
\hline

RX~J0822.0--4300 & $0.112$ & $8.300e-4$ & $$ & $213.799$ & G260.4--3.4 (Puppis A) & $3.700$ & $5.200$ [11] \\
1E~1207.4--5209 & $0.424$ & $6.600e-6$  & $$ & $1.018e5$ & G296.5 +10.0 (PKS 1209--51/52) & $2.000$ & $20.000$ [12] \\
CXOU~J185238.6+004020 & $0.105$ & $8.680e-7$  & $$ & $1.917e5$ & G033.6+00.1 (Kes 79) & $5.400$ & $7.500$ [13] \\ \hline

\end{tabular}
\caption{For a given PSR, $P$ is the period, $\dot{P}$ the period derivative. The characteristic age is $\tau_\text{PSR}$ and the lower and upper SNR age limits are $\tau_\text{SNR--}$ and $\tau_\text{SNR+}$, respectively, from the McGill magnetar catalogue \citep[\protect\url{http://www.physics.mcgill.ca/~pulsar/magnetar/main.html}, ][]{mcGill}. The SNR ages have been compiled in the U. of Manitoba's High-Energy SNR Catalogue (SNRcat, \protect\url{http://www.physics.umanitoba.ca/snr/SNRcat/}). References to SNR ages in this table are [1]: \citet{1841Age}, [2]: \citet{nakano15}, [3]: \citet{1714Age2009}, [4]: \citet{0526Age}, [5]: \citet{1627Age}, [6]: \citet{g292age}, [7]: \citet{moi}, [8]: \citet{1846Age}, [9]: \citet{0537Age}, [10]: \citet{0833Age}, [11]: \citet{0822Age} , [12]: \citet{1207Age}, [13]: \citet{1852Age}. References to the braking indices included here are [14]: \citet{welt11},  [15]: \citet{youngn}, [16]: \citet{1846-3}, [17]: \citet{1846-4}, [18]: \citet{n0537}, [19]: \citet{n0833}.}
\label{table2}
\end{center}
\end{table*}

\begin{center}
\begin{figure*}
\centerline{\includegraphics[scale=1, bb= 61 217 542 580, clip=true]{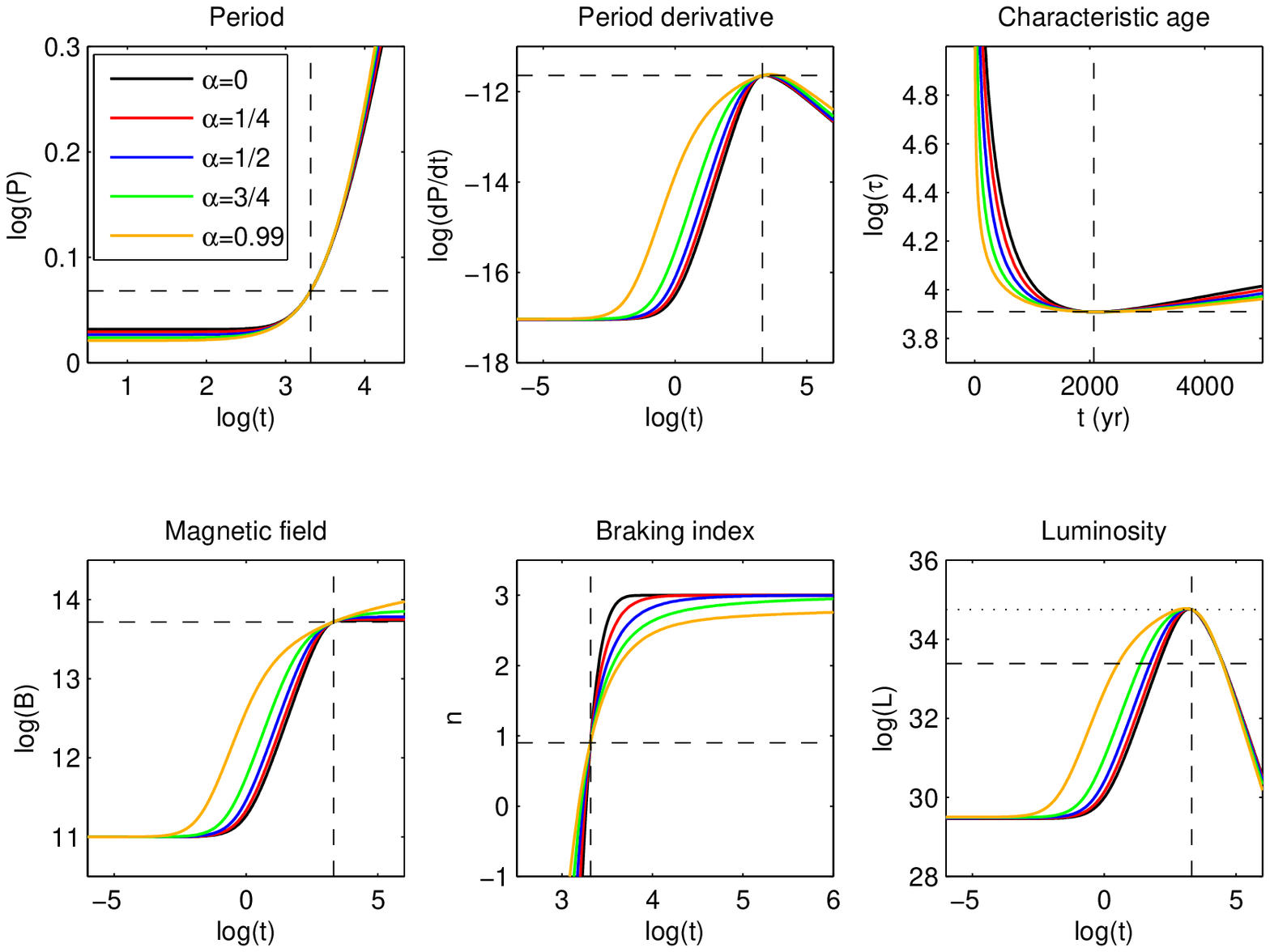}}
\caption{Fits to HBP~J1734--3333 for a variety of field index $\alpha$. The vertical dashed line marks the adopted age $t=2.07$ kyr, chosen to facilitate comparison with the result from \citet{ho15}. The horizontal dashed lines mark the observed quantities. In the luminosity plot (lower right-hand panel) the horizontal dotted line marks the spin-down energy loss rate and the horizontal dashed line marks the observed X-ray luminosity.}
\label{fig:1734}
\end{figure*}
\end{center}

For the purpose of illustrating the effect that changing $\alpha$ has on the field evolution, we consider the HBP~J1734--3333 as an example and use the age derived in \citet{ho15}, $t=2.07$ kyr. We arbitrarily set the initial field to a typical NS field strength, $B_\text{0}=10^{11}$ G, which fixes $\epsilon$ for a given $B_\text{G}$ by equation (\ref{BInit}). We generate a family of curves using a constant field index that spans the full range $0 \leq \alpha<1$. For each value of $\alpha$, we follow the standard model fitting approach, treating the time-scale $\tau_\text{m}$, asymptotic field $B_\text{G}$ and initial period $P_\text{0}$ as fit parameters, which are varied numerically. The results are shown in Fig. \ref{fig:1734}, where we plot the period, period derivative, characteristic age, magnetic field, braking index and luminosity as functions of time for each of the $\alpha$ values, matching to the observed $P$, $\text{d}P/\text{dt}$ and $n$. The horizontal dashed lines represent the observed values and the vertical dashed line is the adopted current age $t$. In the luminosity panel, we show the spin-down luminosity ($\dot{E}$) as a horizontal dotted line and the 2--10 keV X-ray luminosity ($L_\text{x}$) as a dashed line. We call attention to two important features of this figure. First, the characteristic age decays rapidly from an initially high value regardless of $\alpha$. This general behaviour provides an explanation for young NSs that have a characteristic age larger than the corresponding SNR age. Secondly, the large characteristic age at early times gives a negative braking index at early times through equation (\ref{n}), which allows the field growth scenario to explain the observations of objects with $n<0$, such as PSR J0537--6910 with $n=-1.5$ \citep{n0537}. The field index $\alpha$ smoothly controls how quickly the braking index reaches the asymptotic value $n=3$.

Next, we use our field growth model and contour approach to recover the quantitative behaviour of the detailed simulation performed by \citet{ho15}. We assume the value of the asymptotic field and age derived in that work a priori, and then we treat the time-scale $0.1 \leq \tau_\text{m} \leq 10$ kyr and field index $0 \leq \alpha \leq 0.99$ as free parameters. We construct a grid of parameter values ($\tau_\text{m}$, $\alpha$) and calculate $t$ and $n$ for each. Then we find the level sets of these functions at the derived age and mean braking index. The intersection of the two curves gives a unique $\tau_\text{m}$ and $\alpha$ pair. With $\epsilon=0.0021$ we find $P_\text{0}=1.0597$ s, compared to $1.06$ s given by \citet{ho15}. The initial period does not change significantly for lower $\epsilon$. The discrepancy grows slowly as $\epsilon$ is increased. Using the contour approach we find the parameters necessary for our model to reproduce the evolutionary trajectories of PSR~J1734--3333, PSR~B0833--45 and PSR~J0537--6910 from \citet{ho15}. The details of the fits are given in Table \ref{table:fitPar} and marked with an asterisk. It is a testament to the flexibility and usefulness of the parametric form that we were able to recover the behaviour of a simulation involving detailed and complex physical processes.

\begin{center}
\begin{table*}
\begin{tabular}{|c|c|c|c|c|c|c|}
\multicolumn{7}{c}{Fit parameters} \\
\hline
\hline

PSR & $\alpha$ & $\epsilon$ & $\tau_{\text{m}}$ & $B_{\text{G}}$             & $P_0$ & $t$ \\
        &               &                   &    (kyr)          &  ($10^{13}$         G)                    &     (s)      & (kyr) \\ \hline
\hline

SGR~0526--66 & $0.100$ & $0.100$ & $1.642$ & $56.035$ & $3.246$ & $4.786$ \\
                         & $0.900$ & $0.001$ & $0.438$ & $240.000$ & $3.689$ & $4.779$ \\ \hline

SGR~1627--41 & $0.100$ & $0.100$ & $1.261$ & $22.488$ & $0.090$ & $3.677$ \\
		   & $0.900$ & $0.001$ & $3.201$ & $240.000$ & $0.477$ & $4.981$ \\ \hline
\hline

HBP~J1734--3333    & $0.100$ & $0.100$ & $1.193$ & $5.231$ & $1.012$ & $3.477$ \\
*		       	 & $0.633$ & $0.002$ & $0.085$ & $6.500$ & $1.060$ & $2.070$ \\
    			 & $0.900$ & $0.001$ & $10.000$ & $75.236$ & $0.838$ & $9.939$ \\ \hline

HBP~J1846--0258 A & $0.100$ & $0.001$ & $0.074$ & $4.932$ & $0.247$ & $0.433$ \\
			  & $0.900$ & $0.100$ & $6.718$ & $43.273$ & $0.007$ & $0.810$ \\ \hline

HBP~J1846--0258 B & $0.100$ & $0.100$ & $0.286$ & $4.880$ & $0.187$ & $0.833$ \\
	   		 & $0.900$ & $0.100$ & $3.102$ & $43.187$ & $0.225$ & $0.430$ \\ \hline
\hline

PSR~J0537--6910 & $0.100$ & $0.100$ & $0.372$ & $0.094$ & $0.015$ & $1.007$ \\
*		         & $0.525$ & $0.053$ & $0.921$ & $0.170$ & $0.015$ & $1.950$ \\
		         & $0.900$ & $0.001$ & $10.000$ & $2.942$ & $0.014$ & $3.564$ \\ \hline

PSR~B0833--45   & $0.100$ & $0.100$ & $2.254$ & $0.338$ & $0.073$ & $6.570$ \\
*		       & $0.541$ & $0.005$ & $2.709$ & $0.550$ & $0.065$ & $10.200$ \\
		       & $0.900$ & $0.001$ & $10.000$ & $3.586$ & $0.058$ & $15.738$ \\  \hline
\hline

RX~J0822.0--4300 & $0.100$ & $0.100$ & $1.783$ & $0.098$ & $0.111$ & $5.200$ \\
			& $0.900$ & $0.001$ & $10.000$ & $3.010$ & $0.112$ & $3.702$ \\ \hline

1E~1207.4--5200 & $0.100$ & $0.001$ & $6.860$ & $0.017$ & $0.420$ & $20.000$ \\
		        & $0.900$ & $0.001$ & $10.000$ & $0.880$ & $0.424$ & $2.004$ \\ \hline

CXOU~J185238.6+004020 & $0.100$ & $0.100$ & $2.572$ & $0.003$ & $0.105$ & $7.500$ \\
				  & $0.900$ & $0.001$ & $10.000$ & $0.069$ & $0.105$ & $5.401$ \\ \hline
\hline

\end{tabular}
\caption{Fit parameters of the NSs plotted in Figs. \ref{fig:phaseSpace} and \ref{fig:age}.  The solutions with low and high asymptotic fields are listed, and solutions that recover the parameters of \citet{ho15} are marked with an asterisk.}
\label{table:fitPar}
\end{table*}
\end{center}

\begin{center}
\begin{figure*}
\centerline{\includegraphics[scale=0.90, bb= 123 221 476 572, clip=true]{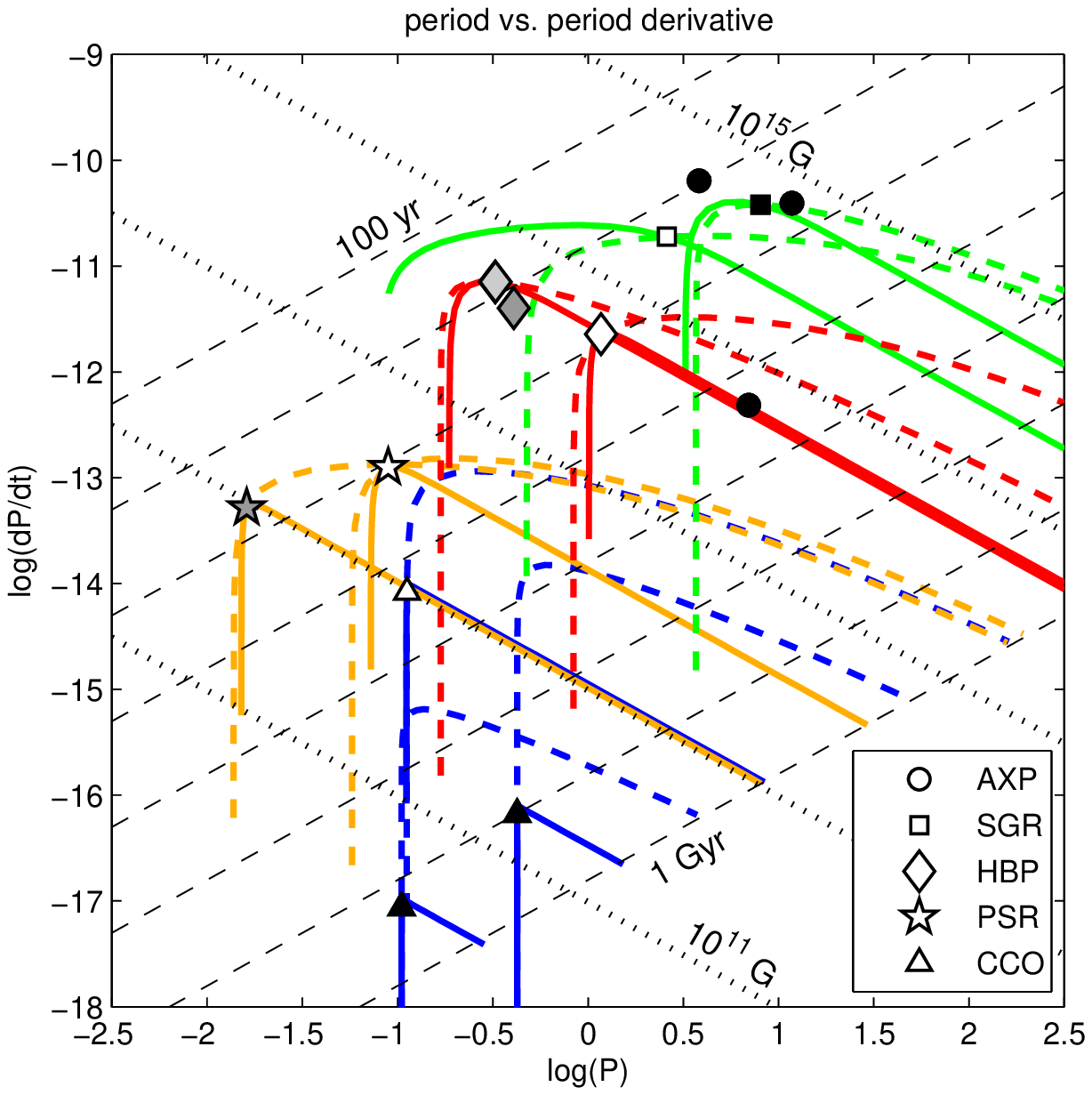}}
\caption{Phase space plot of evolutionary trajectories $P$--$\dot{P}$. The thin black dashed diagonal lines denote constant characteristic age from $100$ yr (upper left) to $1$ Gyr (lower right). The dotted black diagonal lines represent an increasing magnetic dipole field, from $10^{11}$ G (lower left) to $10^{15}$ G (upper right). Black symbols represent sources that have X-ray luminosity in excess of spin down luminosity. The low asymptotic field trajectories are marked as solid lines, and the high field solutions from Table \ref{table:fitPar} are dashed lines. The evolutionary tracks for HBPs are red, SGRs are green, PSRs are yellow and the CCOs denoted with blue. HBP~J1846--0258 is marked by a light grey diamond and the post-outburst trajectory is shown. HBP~J1734--3333 is a white diamond and HBP~J1119--6127 is a dark grey diamond (note that this object is not accompanied with a trajectory). The PSRs J0537--6910 and B0833--45 are marked by grey and white stars.}
\label{fig:phaseSpace}
\end{figure*}
\end{center}

Finally, we apply the contour method to the remaining HBPs and PSRs listed in Table \ref{table:fitPar}, including the braking index when possible. We follow the prescription outlined for contouring in Section \ref{sec:theory}, holding $\alpha$ and $\epsilon$ constant and finding the level sets of $t$ and $n$ as functions of the time-scale $\tau_\text{m}$ and asymptotic field $B_\text{G}$. We do not consider any asymptotic field strength greater than the maximum observed magnetar field, $B_{\text{G}}=2.4\times 10^{15}$ G, of SGR 1806--20 \citep{maxField}. For each system shown in Table \ref{table1}, we provide example solutions with large and small asymptotic fields in Table \ref{table2}, and plot the trajectories of these example solutions in the $P$--$\dot{P}$ phase space in Fig. \ref{fig:phaseSpace}. In this plot, the evolutionary trajectories of the SGRs are given as green lines, the HBPs as red lines, the PSRs yellow lines and the CCOs as blue lines. The HBP~J1846--0258 is marked by a light grey diamond, HBP~J1119--6127 is a dark grey diamond and HBP~J1734--3333 is a white diamond. The PSRs J0537--6910 and B0833--45 are marked by grey and white stars, respectively. Markers that are black represent objects with X-ray luminosity in excess of spin-down luminosity. The parameters that describe the trajectories shown in this figure are likewise given in Table \ref{table:fitPar}. Note that we do not provide an example trajectory for HBP~J1119--6127, which will be discussed in the next section.

\begin{center}
\begin{figure*}
\centerline{\includegraphics[scale=0.90, bb= 130 220 476 575, clip=true]{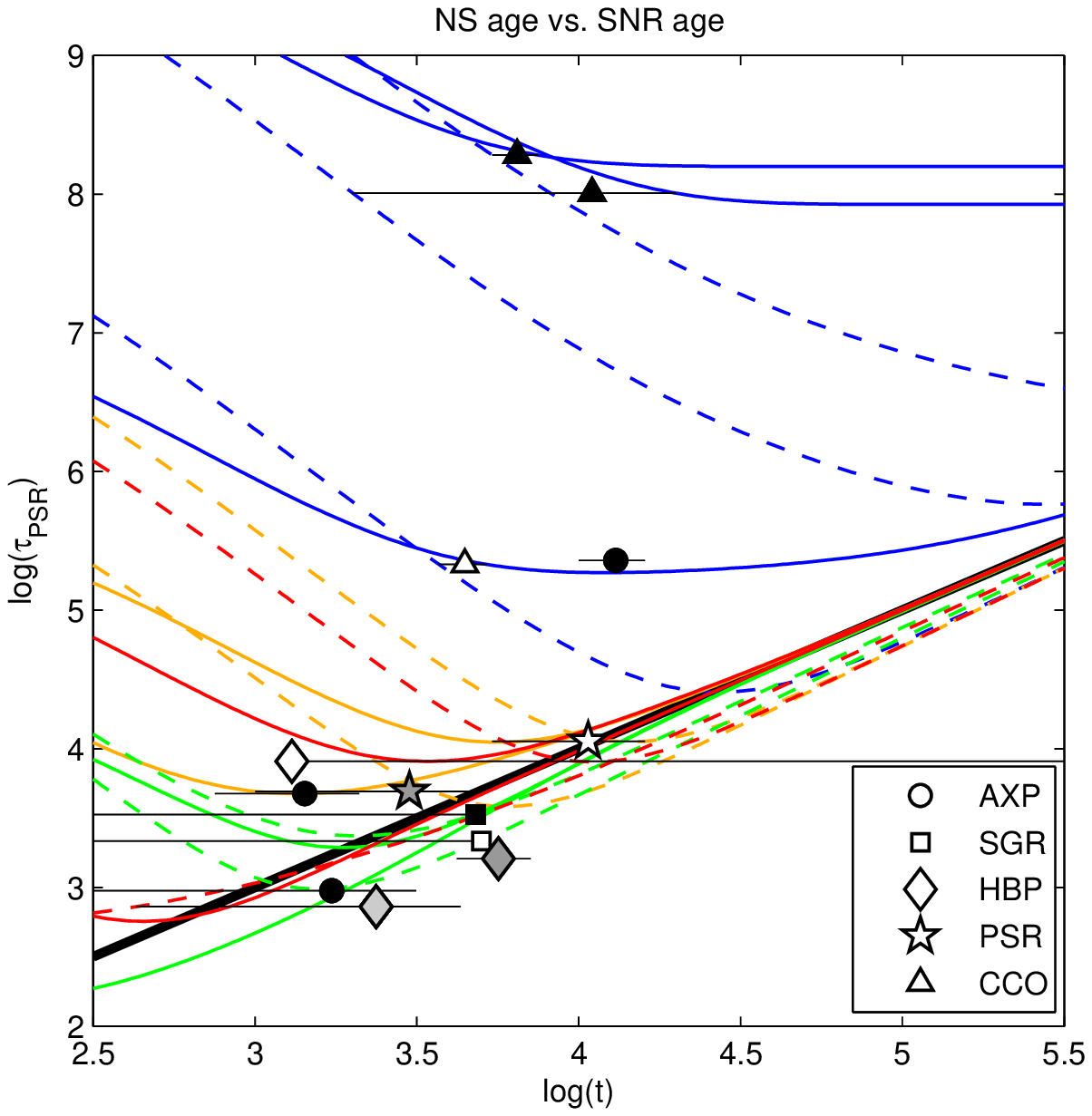}}
\caption{Evolutionary trajectories plotted by NS characteristic age against model time for the systems shown in Table \ref{table:fitPar}. The SNR ages are represented as horizontal lines. The markers are placed at the mean value, or at the extreme when no upper or lower limit exists. The low asymptotic field trajectories are marked as solid lines, and the high field solutions from Table \ref{table:fitPar} are dashed lines. Colours are the same as in Fig. \ref{fig:phaseSpace}. The thick black line is $\tau_\text{PSR}=t$.}
\label{fig:age}
\end{figure*}
\end{center}

In Fig. \ref{fig:age}, we plot the characteristic age against the model time, following the same conventions as Fig. \ref{fig:phaseSpace}. Despite the apparent similarity of many of the trajectories in the $P$--$\dot{P}$ phase space, the $\tau$--$t$ plot clearly shows the difference between these objects as a function of time. It is worth noting that the time evolution of the CCOs characteristic age explains the apparent large discrepancy between the pulsars adopted ages (appearing very old) and their associated young SNRs. In particular, for the three systems shown, the PSR and SNR ages match at times equal to or exceeding $\sim 10^{4.5}$ yr, by which time the SNR would have mostly dissipated. Therefore, the characteristic age for these objects considered will not reflect their true age as long as they are within their SNRs. This feature, along with the low asymptotic field strength (see Table \ref{table:fitPar}), also leads to the suggestion that CCOs could be ancestors of `old' isolated radio pulsars as long as they overcome the accretion or field-growth phase (which would explain their X-ray dominant emission) and their surface field grows to the critical limit required for radio emission. The late time evolution of the CCOs may also link them to the class of objects known as X-ray dim isolated NSs \citep[XDINS; ][]{xdin1}. These are radio-quiet X-ray pulsars with long periods ($3.45$--$11.37$ s) and no apparent SNR associations. Some of these objects are believed to have high magnetic fields in excess of $10^{13}$ G.

\section{Discussion}
\label{sec:discussion}

The $P$-$\dot{P}$ phase space trajectories shown in Fig. \ref{fig:phaseSpace} demonstrate possible evolutionary links between the apparently diverse set of NSs shown in Table \ref{table:fitPar}. As the NS fields grow, they evolve from the bottom of the figure upward, passing through the region of the phase space inhabited by the CCOs. Thus the PSRs J0537--6910 and B0833--45, and the HBPs J1846--0258 and J1734--3333 may have undergone a similar CCO stage during their evolution. The trajectories of these HBPs carry them towards the current position of AXP 1E2259+586. If HBPs and AXPs are related through evolution then field decay must begin once the buried field has emerged, raising the braking index to values $n>3$. We have also fit the CCOs to explore their potential future behaviour, and note that RX~J0822.0--4300 has asymptotic behaviour for both large and small fields which is very close to PSR~B0833--45. However, the time evolution of these objects is dramatically different as seen in Fig. \ref{fig:age}. Thus, objects like the HBPs may pass through the CCO stage relatively quickly, whereas objects such as CCO~1E~1207.4--5209 spend a more significant portion of their lives in this state. Finally, we note that CCO~CXOU~J185238.6+004020 requires an extremely low asymptotic field, with $B_{\text{G}}<6.9 \times 10^{11}$ G. Thus, even after the field emerges from this NS, it remains relatively low.

Since the SGRs 0527--66 and 1627--41 have characteristic ages less than the upper SNR age limit, we have also examined these objects using the growth model. However, the field growth mechanism is not generally expected to play a role in the evolution of the SGRs, since their characteristic ages are only smaller than the upper limit of the associated SNR age, and no lower limits are known. Moreover, these systems do not have a measured braking index, which is crucial in making the case for field growth ($n<3$) or field decay ($n>3$). Additionally, field decay has been proposed to explain the SGRs energetics as it has for AXPs, although their X-ray luminosity is not consistently larger than their spin-down energy. Due to the lack of a lower age limit, we consider solutions that produce $\tau_\text{PSR} < t \leq \tau_{\text{SNR}+}$. Generally solutions that satisfy this condition with large $B_\text{G}$ require a longer growth time-scale for a given $\alpha$ and $\epsilon$ pair, so we can find large fields $B_\text{G} > 2.0\times 10^{15}$ G provided we consider sufficiently large $\tau_\text{m}$. We plot the evolution of two example solutions in Figs \ref{fig:phaseSpace} and \ref{fig:age}. Interestingly, both SGR~1627--41 and SGR~0527--66 reach similar states in the limit of large asymptotic fields shown in Fig. \ref{fig:phaseSpace}, and the trajectories imply the SGR fields are still evolving. A lower limit for the SNR age would significantly constrain these results, provided that $\tau_{\text{SNR}-} > \tau_\text{PSR}$. For completion, we attempted fits to the AXPs as well, but these required unrealistically high initial spin periods. We stress that despite these interesting fits, field decay is necessary to reconcile the characteristic and SNR ages of the AXPs. This conclusion is supported by results from the literature \citep[e.g.][]{nakano15}, and is implied for the SGRs evolution as well \citep{dallOsso}.

HBP~J1846--0258 presents an interesting case since the braking index has been observed to decrease from $n=2.64$ to $2.16$ \citep{1846-3, 1846-4} following an outburst and spectral changes in 2008 \citep{HBP2, HBP1}. This braking index change was not accompanied by a change in luminosity or pulse profile which is difficult to explain on such short time-scales, but may represent a re-organization of the magnetosphere \citep{1846-5}. We fit the pre- and post-outburst configurations of the system which we label as A ($n=2.64$) and B($n=2.16$). However, we were not able to find any region of parameter space through our contour methods that could simultaneously satisfy both pre- and post-outburst configurations. This may not be the case if the field index were allowed to vary in time, but with constant $\alpha$, field growth cannot neatly explain the behaviour of this NS. The HBP~J1846--0258 is a complicated case, particularly because of the presence of a bright pulsar wind nebula powered by this object. This nebula implies wind-braking likely plays an important role in the evolution of this NS. For $\alpha=0.99$ and minimum $\epsilon=0.019$, we find a maximum field $B_\text{G}=6.5 \times 10^{14}$ G on a growth time-scale $9.2$ kyr. Changing $\alpha$ and $\epsilon$ results in a lower field on shorter time-scales. For a given pair of $\alpha$ and $\epsilon$, the system can be well constrained by the period condition and the SNR age, though low remnant ages are generally favoured.

Finally, there was a problem fitting to HBP~J1119--6127. For this system we were not able to simultaneously fit both the age of the associated SNR G292.2--0.5 \citep{g292age}, and the observed braking index $n=2.684$ \citep{welt11}. With the SNR age constraint the derived braking index is $n \approx 1.8$ assuming $t=4.2$ kyr. Alternatively, with the braking index constraint in place, the derived age was found to be close to the characteristic age, and cannot be reconciled with the observed SNR age. Intriguingly, a low value for the braking index of J1119--6127 was also proposed by \citet{g292age}, who suggested that the braking index may have recently changed from a lower value $n < 2$. Since neither of these scenarios satisfies the constraints, we have not included an evolutionary track for HBP~J1119--6127 in Figs \ref{fig:phaseSpace} and \ref{fig:age}, and also exclude it from our summary of solutions in Table \ref{table:fitPar}. We plan to investigate HBP~J1119--6127 with other emission mechanisms in future work.

It is also relevant that many of the systems included in Table \ref{table:fitPar} have large initial spin periods (i.e. $P_0 \approx 1$ s, and approaching $P$ for many systems), which are higher than expected for the traditional magnetar model \citep{magnetar1, IP_1}. This is notable because a problem with the magnetar model is the lack of super energetic SNRs which would be expected from an SNR hosting a rapidly spinning proto-NS \citep[for example,][]{vink, magnetarSNRs}.

\section{Conclusions}
\label{sec:conclusions}

We have devised a flexible and conveniently parametrized model for a growing magnetic field, which is based on the parametric forms used by \citet{colpi} and refined by \citet{dallOsso}. This parametrization can accommodate a variety of field time-dependence in addition to the exponential model suggested by \citet{NB15}, and the interpretation of the parameters is straightforward. By including the observationally measured period and period derivative and assuming the field index and growth parameter $\epsilon$ constant, we are able to study the portions of the parameter space containing solutions which reproduce the observables, without the need for an external optimization routine. We have shown that this phenomenological model is able to reproduce the detailed simulations of field growth by \citet{ho15} to high accuracy. By fitting the HBPs securely associated with SNRs with known ages and measured braking indices, we found interesting evolutionary trajectories for the systems in phase space. We conclude that if field growth is significant in the life cycle of HBPs, then they may be closely related to the CCOs early in their evolutionary histories. The end result of the field growth in CCOs may connect these objects to the HBPs and XDINSs. We also investigate the possibility of field growth in SGRs, however, the behaviour of these systems are largely unconstrained due to the absence of a lower SNR age limit and lack of measured braking index.

Field growth is not applicable to the AXPs, which require field decay to explain the observed difference in PSR and SNR ages, and a growing field is not necessary to explain the SGRs, provided that the characteristic age is larger than the true SNR age. Thus, in the context of magnetic field evolution, we conclude that both field growth and decay processes are required to explain the diverse population of NSs. Once the field has reached its asymptotic value, field decay may begin, increasing the braking index to values $n>3$ later in life. Thus, the time dependence of magnetic fields provides an interesting avenue to unify the population of NSs, and in particular, explain the apparently large characteristic ages for systems associated with relatively young SNRs.

While this evolutionary picture is simple and based on a phenomenological model, there are many emission mechanisms which have been proposed in the literature to solve the braking index and PSR--SNR age discrepancy problems. The field growth model was shown to be ineffective in explaining the constraints present in the HBP~J1119--6127 and the time evolution of HBP~J1846--0258, both of which are associated with pulsar wind nebulae. In a follow-up paper, we will thoroughly investigate alternatives to the physical emission mechanisms at work in these and various other classes of objects and the subsequent implications for the PSR--SNR association and evolution.

\section{Acknowledgements}
This work was primarily supported by the Natural Sciences and Engineering Research Council of Canada (NSERC) through the Canada Research Chairs Programme. SSH is also supported by an NSERC Discovery grant and the Canadian Space Agency. This research made use of NASA's Astrophysics Data System, McGill's magnetars catalogue and the U. of Manitoba's high-energy SNR catalogue (SNRcat). We also thank the referee for the thoughtful comments that helped significantly improve the clarity of this work.

\label{lastpage}

\end{document}